
\input phyzzx

\def\IR{{\hbox{{\rm I}\kern-.2em\hbox{\rm R}}}}
\def\IB{{\hbox{{\rm I}\kern-.2em\hbox{\rm B}}}}
\def\IN{{\hbox{{\rm I}\kern-.2em\hbox{\rm N}}}}
\def\IC{{\ \hbox{{\rm I}\kern-.6em\hbox{\bf C}}}}

\def\IZ{{\hbox{{\rm Z}\kern-.4em\hbox{\rm Z}}}}
\def\to{\rightarrow}

\def\underarrow#1{\vbox{\ialign{##\crcr$\hfil\displaystyle
{#1}\hfil$\crcr\noalign{\kern1pt
\nointerlineskip}$\longrightarrow$\crcr}}}
%

\def\ltorder{\mathrel{\raise.3ex\hbox{$<$}\mkern-14mu
             \lower0.6ex\hbox{$\sim$}}}
\def\lesssim{\mathrel{\raise.3ex\hbox{$<$}\mkern-14mu
             \lower0.6ex\hbox{$\sim$}}}

\def\R{{\bf R}}


\input phyzzx
\overfullrule=0pt
\tolerance=5000
\overfullrule=0pt
\twelvepoint

\twelvepoint
\rightline{CINVESTAV-FIS 22/95}
\date{December, 1995}
\titlepage
\title{ FURTHER REMARKS ON THE CHIRAL MODEL APPROACH TO SELF-DUAL GRAVITY}
\vglue-.25in
\author{Hugo Garc\'{\i}a-Compe\'an, \foot{E-mail: compean@fis.cinvestav.mx}
Jerzy F. Pleba\'nski \foot{E-mail: pleban@fis.cinvestav.mx} and Maciej
Przanowski\foot{Permanent address: Institute of Physics, Technical University
of L\'od\'z, W\'olcza\'nska 21
9, 93-005 L\'od\'z, Poland.}}
\medskip
\address{Departamento de F\'{\i}sica
\break  Centro de Investigaci\'on y de
Estudios Avanzados del IPN.
\break Apdo. Postal 14-740, 07000, M\'exico D.F., M\'exico.}
\bigskip
\abstract{ It is shown how some results on harmonic maps within the chiral
model can be carried over to self-dual gravity. The WZW-like action for
self-dual gravity is found.}


\endpage

\chapter{Introduction}

\REF\W{ R.S. Ward, Class. Quantum Grav. 7 (1990) L217.}

\REF\Husain{V. Husain, Phys. Rev. Lett. 72 (1994) 800; Class. Quantum Grav. 11
(1994) 927.}

\REF\Pleb{J.F. Pleba\'nski, M. Przanowski and H. Garc\'{\i}a-Compe\'an,
 From Principal Chiral Model to Self-dual Gravity, submitted to Mod. Phys.
Lett. A.}

\REF\Fairlie{D.B. Fairlie, P. Fletcher and C.K. Zachos, J. Math. Phys. 31
(1990) 1088.}

\REF\Uhlen{ K. Uhlenbeck, J. Diff. Geom. 30 (1989) 1.}

\REF\Ward{ R.S. Ward, Commun. Math. Phys. 123 (1990) 319.}

\REF\Ana{C.K. Anand, Uniton Bundles, e-print archive dg-ga/9508011 (1995).}

\REF\Comp{H. Garc\'{\i}a-Compe\'an and T. Matos, Phys. Rev. D52 (1995) 4425.}

\REF\Wess{ J. Wess and B. Zumino, Phys. Lett. B37 (1971) 95.}

\REF\Witten{E. Witten, Nucl. Phys. B223 (1983) 422; Commun. Math. Phys. 92
(1984) 455; Commun. Math. Phys. 144 (1992) 189.}

\REF\Stra{I.A.B. Strachan, Phys. Lett. B283 (1992) 63.}

\REF\Boyer{ C.P. Boyer, J.D. Finley III and J.F. Pleba\'nski, Complex general
relativity, ${\cal H}$ and ${\cal HH}$ spaces-a survey, in General relativity
and gravitation, Einstein memorial volume, v.2, ed. A. Held (Plenum, New York,
1980) pp. 241-281.}

\REF\Po{K. Pohlmeyer, Commun. Math. Phys. 46 (1976) 207.}

Some years ago Ward [\W] and Husain [\Husain] showed that self-dual gravity
could be considered as a principal chiral model within the su$(\infty)$ or the
Poisson algebras.

In our previous paper [\Pleb] the analogous point of view has been presented
but we have considered self-dual gravity to be the $\hbar \to 0 $ limit of the
principal chiral model in the Moyal bracket algebra. This last approach seems
to be natural and con
venient as the associative $*$-algebra defining the Moyal bracket algebra has a
transparent interpretation.

 I.A.B. Strachan and T. Matos suggested (private communication) that the
results of [\Pleb] should be very closed related to the considerations of
Fairlie, Fletcher and Zachos [\Fairlie] on the su$(N)$ algebra as embedded in
the Moyal bracket algebra and
also to some results on the harmonic maps, especially to ones given by
Uhlenbeck [\Uhlen] (see also [\Ward,\Ana,\Comp]).

The aim of the present paper is to give a partial answer to those suggestions.

In Section 2 we recall the main points of the previous work [\Comp]. In Section
3, using the embedding of su$(N)$ in the Moyal bracket algebra [\Fairlie], the
construction of solutions to Husain's heavenly equation [\Husain] is given
which, in fact, appea
rs to be the Fourier expansion with the coefficients satisfying SU$(N)$
principal chiral equations for $N \to \infty$.

Section 4 is devoted to indicating how Uhlenbeck's considerations on harmonic
maps into finite Lie groups [\Uhlen] can be carried over to the case of the
groups defined by the Moyal $*$-product. Finally, in Section 5 we consider the
Wess-Zumino-Witten-lik
e (WZW-like) action [\Wess,\Witten] within the Moyal formalism and it is shown
that the action for Husain's heavenly equation is the $\hbar \to 0$ {\it limit
of the WZW-like action in the Moyal algebra}.

\chapter{The Moyal deformation of Husain's heavenly equation}

In this section we recall some results of our previous paper [\Pleb].

Let ${\bf G}$ be a matrix Lie group and ${\cal G}$ its Lie algebra. Consider
the ${\bf G}$ principal chiral model in a connected and simply connected
submanifold $\Omega$ of the two-dimensional Euclidean space $\Re^2$.

The principal chiral equations read

$$\partial_x A_y - \partial_y A_x + [A_x,A_y] = 0, \eqno(2.1a)$$
$$ \partial_x A_x + \partial_y A_y = 0, \eqno(2.1b)$$
where $A_{\mu} \in {\cal G} \otimes C^{\infty}({\Omega})$, $\mu \in \{x,y\}$,
stand for the chiral potentials and $(x,y)$ are the Cartesian coordinates in
$\Omega$. Then, from $(2.1b)$ it follows that there exists a function $\theta =
\theta (x,y) \in {\cal G} \otimes C^{\infty}(\Omega)$ such that

$$ A_x = - \partial_y \theta, \ \ \ \ \ {\rm and} \ \ \ \ \  A_y =
\partial_x \theta. \eqno(2.2)$$
Inserting $(2.2)$ into $(2.1a)$ one gets the principal chiral equations to read

$$ \partial_x^2 \theta + \partial_y^2 \theta + [\partial_x \theta, \partial_y
\theta] = 0. \eqno(2.3)$$
Equivalently, we can proceed as follows. From $(2.1a)$ it follows that
$A_{\mu}$, ${\mu} \in \{x,y\},$ is of the pure gauge form, {\it i.e.}, there
exists a ${\bf G}$-valued function $g = g (x,y)$ such that

$$ A_{\mu} = g^{-1} \partial_{\mu} g. \eqno(2.4)$$
Substituting $(2.4)$ into $(2.1b)$ we get another form of the principal chiral
equations
$$ \partial_{\mu} \big( g^{-1} \partial_{\mu} g \big) = 0 \eqno(2.5)$$
(Summation over $\mu$ is assumed.).

Eq. (2.5) means that $g : \Omega \to {\bf G}$ {\it is a harmonic map from } $
\Omega \subset \Re^2$ {\it into the Lie group} ${\bf G}$ (or, simply, $g:
\Omega \to {\bf G}$ {\it is a  {\bf G} harmonic map}).

Comparing (2.2) and (2.4) one quickly finds the relations

$$ g^{-1} \partial_x g = -\partial_y \theta \ \ \ \ \ \ {\rm and} \ \ \ \ \
g^{-1} \partial_y g = \partial_x \theta. \eqno(2.6)$$

Then, the Lagrangians leading to (2.3) or (2.5) read (under the assumption that
the algebra ${\cal G}$ is semisimple)

$$ {\cal L}_{Ch} := \alpha {\rm Tr} \bigg \{ {1\over 3} \theta [\partial_x
\theta,
\partial_y \theta] - {1 \over 2} \bigg( (\partial_x \theta)^2 + (\partial_y
\theta)^2 \bigg) \bigg \}  \eqno(2.7)$$
or

$${\cal L}'_{Ch}= - \beta {\rm Tr} \big \{ (g^{-1} \partial_{\mu}
g)(g^{-1}\partial_{\mu} g) \big \}$$
$$ = \beta {\rm Tr} \big \{ (\partial_{\mu} g)(\partial_{\mu}g^{-1}) \big \},
\eqno(2.8)$$
respectively; where $\alpha > 0$ and $\beta > 0$ are some constants.

Now, let $ \hat {\bf G}$ be some Lie group of linear operators acting on the
Hilbert space $L^2(\Re^1)$ and let $\hat{\cal G}$ be  the Lie algebra of  $
\hat {\bf G}$. Consider the $ \hat {\bf G}$ principal chiral model. The
principal chiral equations rea
d now

$$\partial_x \hat{A}_y - \partial_y \hat{A}_x + [\hat{A}_x,\hat{A}_y] =
0, \eqno(2.9a)$$
$$ \partial_x \hat{A}_x + \partial_y \hat{A}_y = 0, \eqno(2.9b)$$
where $\hat{A}_{\mu} = \hat{A}_{\mu}(x,y) \in \hat{\cal G} \otimes C^{\infty}
({\Omega}),  \ \mu \in \{x,y\}.$

Analogously as before, from $(2.9b)$ it follows that there exists
$\hat{\theta} = \hat {\theta}(x,y) \in \hat {\cal G} \otimes C^{\infty}
(\Omega)$ such that

$$ \hat{A}_x = - \partial_y \hat{\theta}, \ \ \ \ \ {\rm and} \ \ \ \ \
\hat{A}_y = \partial_x \hat{\theta}. \eqno(2.10)$$

Then, $(2.9a)$ under (2.10) gives

$$ \partial^2_x \hat \theta + \partial^2_y \hat \theta + [\partial_x \hat
\theta, \partial_y \hat \theta] = 0. \eqno(2.11)$$

Equivalently, from $(2.9a)$ one infers that
$$ \hat{A}_{\mu} = \hat{g}^{-1} \partial_{\mu} \hat{g}, \ \ \ \   \mu \in
\{x,y\}, \eqno(2.12)$$
where $\hat{g} = \hat{g}(x,y)$ is some $\hat{\bf G}$-valued function on
$\Omega$. Substituting (2.12) into $(2.9b)$ we get

$$ \partial_{\mu} \big( \hat{g}^{-1} \partial_{\mu} \hat{g} \big) = 0.
\eqno(2.13)$$

It is convenient to define a new operator-valued function $\hat{\Theta} =
\hat{\Theta}(x,y)$ by

$$ \hat{\Theta} := i \hbar \hat{\theta}. \eqno(2.14)$$

Thus, by (2.11),  $\hat{\Theta}$ satisfies the following equation

$$ \partial^2_x \hat \Theta + \partial^2_y \hat \Theta + {1 \over i{\hbar}}
[\partial_x \hat \Theta, \partial_y \hat \Theta] = 0. \eqno(2.15)$$

\noindent
The {\it Weyl correspondence} ${\cal W}^{-1}$ leads from $\hat{\Theta} = \hat
{\Theta}(x,y)$ to the function  $\Theta = \Theta (x,y,p,q)$,  $ \Theta  = {\cal
W}^{-1}(\hat{\Theta})$, defined on $\Omega \times \Re^2$ according to the
formula

$$ \Theta = \Theta(x,y,p,q) := \int_{- \infty}^{+ \infty} {\rm exp} \big( {ip
\xi \over {\hbar}} \big) < q- {\xi\over 2}| \hat {\Theta}(x,y)| q + {\xi \over
2}> d\xi. \eqno(2.16)$$

Then $\Theta$ satisfies the {\it Moyal deformation of Husain's heavenly
equation}

$$ \partial^2_x \Theta + \partial^2_y \Theta + \{ \partial_x \Theta, \partial_y
\Theta \}_M = 0, \eqno(2.17)$$
where $\{\cdot,\cdot \}_M$ denotes the {\it Moyal bracket} {\it i.e.},

$$ \{f_1,f_2\}_M := {1 \over i {\hbar}} (f_1 * f_2 - f_2* f_1) = f_1 {2 \over
{\hbar}} {\rm sin}({{\hbar}\over 2} \buildrel {\leftrightarrow} \over {\cal P})
f_2, $$

$$\buildrel {\leftrightarrow}\over {\cal P} := {\buildrel
{\leftarrow}\over{\partial} \over \partial q} {\vec  \partial \over \partial p}
- {\buildrel{\leftarrow}\over{\partial} \over \partial p} {\vec \partial \over
\partial q}; \ \ \ f_1=f_1(x,y,p,q),
\ \ \ f_2 = f_2(x,y,p,q). \eqno(2.18)$$

The {\it Moyal $*$-product} is defined by

$$ f_1*f_2 := f_1 {\rm exp} ({i {\hbar} \over 2}
\buildrel{\leftrightarrow}\over{\cal P})f_2. \eqno(2.19)$$
If the functions $f_1$ and $f_2$ are independent of $\hbar$, then

$$\lim_{{\hbar}\to 0} f_1 * f_2 = f_1f_2, \ \ \ \ \   \lim_{{\hbar} \to 0}
\{f_1,f_2\}_M = \{f_1,f_2\}_P := f_1 \buildrel {\leftrightarrow}\over {\cal P}
f_2, \eqno(2.20)$$
where $\{\cdot,\cdot \}_P$ denotes the Poisson bracket.

Equivalently, defining the function

$$ g = g(x,y,p,q) := {\cal W}^{-1}\big( \hat{g}(x,y)\big) \eqno(2.21)$$
one gets from (2.13)

$$ \partial_{\mu} \bigg( g^{- \buildrel{*}\over{1}} * \partial_{\mu} g \bigg) =
0, \eqno(2.22)$$
where $g^{- \buildrel{*}\over{1}}$ denotes the inverse of $g$ in the sense of
the Moyal $*$-product {\it i.e.},

$$ g^{- \buildrel{*} \over{1}} * g = g * g^{- \buildrel{*}\over{1}} = 1.
\eqno(2.23)$$

Comparing (2.22) with (2.5) we can say that the function $g = {\cal
W}^{-1}\big( \hat{g}(x,y)\big)$ defines a {\it harmonic map from $\Omega$ into
the Lie group} ${\bf G}_* := {\cal W}^{-1}(\hat {\bf G})$.

As it has been shown in [\Pleb] the Lagrangians leading to Eq. (2.17) or
Eq. (2.22) read

$$ {\cal L}^{(M)}_{SG} = - {1 \over 3} \Theta * \{\partial_x \Theta, \partial_y
\Theta \}_M + {1 \over 2} \bigg( (\partial_x \Theta)*(\partial_x \Theta) +
(\partial_y \Theta)*(\partial_y \Theta) \bigg) \eqno(2.24)$$
or

$${\cal L}^{\prime (M)}_{SG} := - {{\hbar}^2 \over 2} \big(g^{- \buildrel{*}
\over{1}} * \partial_{\mu} g \big)* \big( g^{- \buildrel {*} \over{1}} *
\partial_{\mu} g \big), \eqno(2.25)$$
respectively. To derive Eq. (2.22) one can use ${\cal L}^{\prime (M)}_{SG}$ or
equivalently, the Lagrangian ${\cal L}^{\prime \prime (M)}_{SG}$ defined by

$$ {\cal L}^{\prime \prime (M)}_{SG} = {{\hbar}^2 \over 2} \big(\partial_{\mu}
g
\big)*\big(\partial_{\mu} g^{-\buildrel{*} \over{1}} \big). \eqno(2.26)$$
(Compare with (2.8)).

Note also the relations which follow from (2.10), (2.12), (2.14) and from the
definitions of $\Theta$ and $g$. Namely

$$\partial_y \Theta = -i \hbar g^{- \buildrel{*}\over{1}} * \partial_x g \ \ \
\ \ \  {\rm and} \ \ \
\ \ \partial_x \Theta = i \hbar g^{-\buildrel{*}\over{1}} * \partial_y g.
\eqno(2.27)$$

Assume now that the function $\Theta$ is analytic in $\hbar$, {\it i.e.,}
[\Stra]

$$ \Theta = \sum_{n = 0}^{\infty} {\hbar}^n \Theta_n, \eqno(2.28)$$
where $\Theta_n = \Theta_n(x,y,p,q),$  $n=0,1,...,$ are independent of $\hbar$.
If $\Theta$ is a solution of (2.17), then by (2.20) one concludes that the
function $\Theta_0$ satisfies {\it Husain's heavenly equation} [\Husain]

$$ \partial^2_x \Theta_0 + \partial^2_y  \Theta_0 + \{ \partial_x  \Theta_0,
\partial_y \Theta_0 \}_P = 0. \eqno(2.29)$$
Moreover, the Lagrangian ${\cal L}_{SG}$ leading to Eq. (2.29) can be quickly
found to read

$${\cal L}_{SG} = \lim_{\hbar \to 0} {\cal L}^{(M)}_{SG} = - {1 \over 3}
\Theta_0 \{ \partial_x \Theta_0, \partial_y \Theta_0 \}_P + {1\over 2} \bigg(
(\partial_x \Theta_0)^2 + (\partial_y \Theta_0)^2 \bigg). \eqno(2.30)$$

\noindent
(The Lagrangians of this type were first considered by Boyer, Finley and one of
us (J.F.P.) [\Boyer] in the context of the first and second heavenly
equations.).

Therefore, {\it self-dual gravity appears to be the $\hbar \to 0$ limit of the
principal chiral model for the Moyal bracket algebra.}

One can interpret self-dual gravity to be the {\it principal chiral model for
the Poisson bracket algebra.} [\W,\Husain].

These two approaches are evidently equivalent. However, the first one seems to
be much more convenient as the Moyal bracket is defined in a natural manner as
the commutator in the associative algebra defined by the Moyal $*$-product
(see(2.18) and (2.19))
. Thus, in a sense the Moyal bracket algebra can be considered to be an
infinite dimensional matrix Lie algebra. Especially interesting is the case
when the group ${\bf G}_*$ is a subgroup of ${\bf U}_*$, where

$$ {\bf U}_* := \{ f = f(p,q) \in C^{\infty}(\Re^2); \ f* \bar{f} = \bar{f} * f
= 1\};\eqno(2.31)$$
(the bar stands for the complex conjugation.).

It means that $\hat{\bf G} = {\cal W}({\bf G}_*)$ is a subgroup of the group
$\hat{\bf U}$    of unitary operators acting on $L^2(\Re^1).$

Now one  quickly finds that if $g = g(x,y,p,q)$ is an ${\bf U}_*$-valued
function, then $ g^{- \buildrel{*}\over{1}} * \partial_{\mu} g = \bar{g} *
\partial_{\mu} g$ is {\it pure imaginary}, and consequently, by (2.27), the
function $\Theta$ can be chosen
 so as to be {\it real}.

Finally, we recall the method proposed in [\Pleb] of searching for solutions of
Husain's heavenly equation (2.29).

Let $\Psi : {\cal G} \to \hat{\cal G}$ be a Lie algebra homomorphism, and let
(summation over $a!$)

$$\theta = \theta (x,y) = \theta_a(x,y) \tau_a \in {\cal G} \otimes
C^{\infty}(\Omega) \eqno(2.32)$$
be some solution of the principal chiral equation (2.3); $\tau_a \in {\cal G}$,
 $a=1,2,..,{\rm dim} \ {\cal G}$ constitute a basis of ${\cal G}$. Then

$$ \hat{\Theta} = \hat{\Theta}(x,y) = i \hbar \theta_a(x,y) \hat{X}_a$$

$$ \hat{X}_a := \Psi(\tau_a) \eqno(2.33) $$
satisfies Eq. (2.15). Therefore, the function $\Theta$ defined by (see (2.16))

$$ \Theta = \Theta(x,y,p,q) = i \hbar \theta_a(x,y) X_a(p,q) $$
$$ X_a(p,q):= {\cal W}^{-1}(\hat{X}_a) \eqno(2.34)$$
fulfills the {\it Moyal deformation of Husain's heavenly equation}, (2.17).

Consequently, if $\Theta$ is of the form (2.28) then $\Theta_0$ satisfies {\it
Husain's heavenly equation}, (2.29). Moreover, if the Lie group $\hat{\bf G}$
defined by the Lie algebra $\hat{\cal G}$ appears to be a subgroup of the group
$\hat{\bf U}$ of
unitary operators in $L^2(\Re^1)$ then the functions $\Theta$ and $\Theta_0$
are real.

\noindent
[Remark:

In contrary to the case of finite matrix groups the equivalence between (2.11)
and (2.13), or (2.17) and (2.22) are not proved to hold in all generality.

We have the implications

$$ (2.13) \Rightarrow (2.11) \ \ \ \ \  {\rm and} \ \ \ \ \  (2.22) \Rightarrow
 (2.17),$$
but the inverse implications, in general, may not be true.

This problem should be considered elsewhere].

\chapter{The construction of solutions of Husain's heavenly equation from
$SU(N)$ chiral fields}

Consider the SU$(N)$ principal chiral model. In this case $A_{\mu} =
A_{\mu}(x,y) \in {\rm su}(N) \otimes C^{\infty}(\Omega),$  $\mu \in \{x,y\}$,
and consequently $\theta = \theta (x,y)$ defined by (2.2) belongs to ${\rm
su}(N) \otimes C^{\infty}(\Omega)
.$

In a distinguished paper by Fairlie, Fletcher and Zachos [\Fairlie] the basis
of the Lie algebra su$(N)$ is defined which appears to be very useful in our
further considerations.

The elements of this basis are denoted by $L_{\bf m}$, $L_{\bf n}$,..., etc.,
${\bf m} = (m_1,m_2)$, ${\bf n} = (n_1,n_2)$,..., etc., and ${\bf m}, \ {\bf
n},...  \in I_N \subset {\bf Z} \times {\bf Z} - \{(0,0) \ {\rm mod} \  N_{\bf
q} \}$
where ${\bf q}$ is any element of ${\bf Z} \times {\bf Z}$. The basic vectors
$L_{\bf m}, \ {\bf m}\in I_N,$ are the $N \times N$ matrices satisfying the
following commutation relations

$$ [L_{\bf m}, L_{\bf n}] = {N \over \pi} {\rm sin} \big( {\pi \over N} {\bf m}
\times {\bf n} \big) L_{{\bf m} + {\bf n} \ \  {\rm mod} N_{\bf q}},
\eqno(3.1)$$
where ${\bf m} \times {\bf n} := m_1n_2 - m_2 n_1$.

Then any solution of the SU$(N)$ principal chiral equations (2.3) can be
written in the form

$$\theta = \theta(x,y) = \sum_{{\bf m} \in I_N} \upsilon_{\bf m}(N;x,y) L_{\bf
m}. \eqno(3.2)$$

Now we let $N$ tend to infinity. In this case  $I_{\infty} \equiv I = {\bf Z}
\times {\bf Z} - \{(0,0)\}$ and the commutation relations (3.1) read

$$[L_{\bf m}, L_{\bf n}] = {\bf m} \times {\bf n} \  L_{{\bf m} + {\bf n}}.
\eqno(3.3)$$

Consider the set $ \{e_{\bf m}\}_{{\bf m} \in I}$, $e_{\bf m} := {\rm exp}
\big[ i(m_1 p + m_2 q) \big]$. One quickly finds that

$$ \{e_{\bf m}, e_{\bf n} \} _P = {\bf m} \times {\bf n} \  e_{{\bf m} + {\bf
n}}. \eqno(3.4)$$

Thus the mapping $F: L_{\bf m} \mapsto e_{\bf m}, \ \  {\bf m} \in I$,  defines
the isomorphism

$$ {\rm su}(\infty) \cong {\rm the \ Poisson \ algebra \ on} \ T^2 \cong {\rm
sdiff}(T^2), \eqno(3.5)$$
where $T^2$ is the 2-torus.

Let now there exists the limit

$$ \upsilon_{\bf m} = \upsilon_{\bf m}(x,y) : = \lim_{N\to {\infty}}
\upsilon_{\bf m} (N;x,y) \eqno(3.6)$$
for every ${\bf m} \in I$. Then it is evident that the function

$$ \Theta_0 = \Theta_0(x,y,p,q) = \sum_{{\bf m} \in I} \upsilon_{\bf m}(x,y) \
{\rm exp} \big[ i(m_1 p + m_2 q) \big] \eqno(3.7)$$
is a solution of Husain's heavenly equation (2.29). The self-dual vacuum metric
$ds^2$ is determined by $\Theta_0$ as follows [\Husain]

$$ ds^2 = dx ( \Theta_{0,xp} dp + \Theta_{0,xq} dq) + dy (\Theta_{0,yp} dp +
\Theta_{0,yq} dq) + {1 \over \{\Theta_{0,x}, \Theta_{0,y} \}_P} \cdot
\big[(\Theta_{0,xp} dp + \Theta_{0,xq} dq )^2 $$

$$+ (\Theta_{0,yp} dp + \Theta_{0,yq} dq)^2 \big], \eqno(3.8)$$
where $\Theta_{0,xp} \equiv \partial_p \partial_x \Theta_0$, $\Theta_{0,yp}
\equiv \partial_p \partial_y \Theta_0,$ ..., etc.

The method presented here leads to the {\it construction of the self-dual
vacuum metrics from the solutions of the} SU{\it (N) principal chiral
equations}. (Compare this with Ward's question: ``Can one construct a sequence
of SU$(N)$ chiral fields, for $N=2,3,...,$ tending to a curved space in the
limit?'' [\W].).

\chapter{ The gravitational uniton}

The aim of the present section is to show how the considerations concerning the
harmonics maps into the Lie group U$(N)$ given in an excellent paper by
Uhlenbeck [\Uhlen] ( See also  Pohlmeyer [\Po], Ward [\Ward] and Anand [\Ana])
can be carry over to sel
f-dual gravity. Our results are very far from that to be complete;
nevertheless, we intend to indicate the direction of further investigations.
First, one can find the Lax pair for the Moyal deformation of Husain's heavenly
equation, (2.17). To this end,
by the analogy to [\Uhlen-\Ana], we consider the following system of linear
partial differential equations on the function $ E = E (\lambda,x,y,p,q) \in
C^{\infty}({\bf C}^* \times \Omega \times \Sigma)$, ${\bf C}^* := {\bf C} -
\{0\}$, $\Sigma \subset \R
e^2$,

$$ i \hbar \partial_{\bar z} E_{\lambda} = ( 1 - \lambda) E_{\lambda} * a_{\bar
z}, $$

$$ i \hbar \partial_z E_{\lambda} = ( 1 - \lambda^{-1}) E_{\lambda} * a_z,
\eqno(4.1)$$
where $\lambda \in {\bf C}^*$, $E_{\lambda} = E_{\lambda}(x,y,p,q) :=
E(\lambda,x,y,p,q)$ and $ z:= x + i y$, $\bar z := x - i y$, $ \partial_z = {1
\over 2} ( \partial_x - i \partial_y)$,  $ \partial_{\bar z} = {1 \over 2} (
\partial_x + i \partial_y)$;
moreover $a_z = a_z(x,y,p,q)$ and  $a_{\bar z} = a_{\bar z}(x,y,p,q)$ are
independent of $\lambda$.

The integrability conditions of the system (4.1) read

$$\lambda^0 \rightarrow  \partial_z a_{\bar z} - \partial_{\bar z} a_z + 2 \{
a_z, a_{\bar z} \}_M = 0 \eqno(4.2a)$$

$$\lambda^1 \rightarrow  \partial_z a_{\bar z} + \{ a_z, a_{\bar z} \}_M = 0
\eqno(4.2b)$$

$$\lambda^{-1} \rightarrow   \partial_{\bar z} a_z +  \{a_{\bar z}, a_z \}_M =
0. \eqno(4.2c)$$
One quickly finds that any two of the above conditions imply the third.

Adding $(4.2b)$ and $(4.2c)$ we get

$$ \partial_z a_{\bar z} + \partial_{\bar z} a_z = 0. \eqno(4.3)$$
Hence, there exists a function $\Theta = \Theta(x,y,p,q)$ such that

$$ a_z = - {i \over 2} \partial_z \Theta \ \ \ \ \  {\rm and} \ \ \ \ \ \
a_{\bar z} = {i \over 2} \partial_{\bar z} \Theta. \eqno(4.4)$$
Substraction  of $(4.2b)$ and $(4.2c)$ gives $(4.2a)$, and with the use of
(4.4) we get finally

$$ i \partial_z \partial_{\bar z} \Theta + {1 \over 2} \{ \partial_z \Theta,
\partial_{\bar z} \Theta \}_M = 0.  \eqno(4.5)$$

One can easily find that Eq. (4.5) written in terms of the coordinates $(x,y)$
is exactly the Moyal deformation of Husain's heavenly equation, (2.17).

Concluding: {\it the system (4.1) constitutes the Lax pair for Eq. (2.17).}

Comparing (4.4) with (2.27) we obtain

$$ a_z = {i \hbar \over 2}  g^{- \buildrel{*}\over{1}} * \partial_z g \ \ \ \ \
\  {\rm and} \ \ \
\ \  a_{\bar z} = {i \hbar\over 2} g^{-\buildrel{*}\over{1}} * \partial_{\bar
z} g, \eqno(4.6)$$
where $g = g(x,y,p,q)$ is a {\it harmonic map from $\Omega$ into
$C^{\infty}(\Sigma)$} {\it i.e.}, the equation

$$ \partial_z \big(  g^{- \buildrel{*}\over{1}} * \partial_{\bar z} g \big) +
\partial_{\bar z} \big( g^{-\buildrel{*}\over{1}} * \partial_z g  \big) = 0
\eqno(4.7)$$
or equivalently Eq. (2.22), hold.

In what follows, if $g=g(x,y,p,q) \in C^{\infty}(\Omega \times \Sigma)$
satisfies Eq. (4.7) we will say that {\it $g$ is a $*$-harmonic map from
$\Omega$ into $C^{\infty}(\Sigma)$.} In particular, consider the case when

$$ g^{- \buildrel{*}\over{1}} = \bar g. \eqno(4.8)$$
In this case we say that {\it $g$ is $*$-unitary} and the group of $*$-unitary
functions will be denoted by  ${\bf U}_*(\Omega \times \Sigma)$ (compare with
(2.31).). Define $a_x = a_x(x,y,p,q)$ and $a_y=a_y(x,y,p,q)$ in a natural
manner

$$ a_z =: {1 \over 2}(a_x - i a_y) \ \ \ \ \ \ \  {\rm and} \ \ \ \ \ \ \
a_{\bar z} =: {1\over 2}(a_x + i a_y). \eqno(4.9)$$
Then by (4.4) we get

$$a_x = - {1\over 2} \partial_y \Theta \ \ \ \ \ \ \  {\rm and} \ \ \ \ \ \ \
a_y = {1\over 2} \partial_x \Theta,  \eqno(4.10)$$
and by (4.6)

$$a_x = {i \hbar \over 2} g^{- \buildrel{*}\over{1}}* \partial_x g \ \ \ \ {\rm
and} \ \ \ \  a_y = {i \hbar \over 2} g^{- \buildrel{*}\over{1}}* \partial_y g.
\eqno(4.11)$$
Therefore, if $g$ is $*$-unitary {\it i.e.} (4.8) holds, then from (4.11) one
infers that $a_x$ and $a_y$ are {\it real} functions. Consequently, by (4.10)
we can choose $\Theta$ to be {\it real}. Then, in the present case

$$ \overline{a_z} = a_{\bar z}. \eqno(4.12)$$

Now we can prove the theorem which is an analogue of the Theorem 2.2 given in
Uhlenbeck's paper [\Uhlen] ( see also [\Ward,\Ana]).

\subsection{ Theorem 4.1}

Let $g=g(x,y,p,q)$ be a $*$-harmonic map from $\Omega$ into
$C^{\infty}(\Sigma)$   and let $g(x_0,y_0,p,,q) = h(p,q),$ where $h=h(p,q)$ is
some given function on $\Sigma$ and $(x_0,y_0)\in \Omega$ is a given point.
Moreover, let for every $\lambda \in {\b
f C}^*$ there exists a unique solution $E_{\lambda} = E_{\lambda}(x,y,p,q)$ of
the Lax pair (4.1) with (4.6), such that $E_{\lambda}(x_0,y_0,p,q)=h(p,q)$ and
$E_{\lambda}$ has the $*$-inverse $E_{\lambda}^{-\buildrel{*}\over{1}}$.

\noindent
Then
\item{{\it (i)}} $E_1 = h(p,q)$,

\item{{\it (ii)}} $E_{-1} = g$,

\item{{\it (iii)}} If $g\in {\bf U}_*(\Omega \times \Sigma)$ then

$$ E_{\lambda}^{-\buildrel{*}\over{1}} = \overline{E_{\bar \lambda^{-1}}} \ \ \
\  {\rm for \ every} \ \  \lambda\in {\bf C}^*. \eqno(4.13)$$

\noindent
{\it Proof}:

The proof of {\it (i)} and {\it (ii)} is trivial.

To prove {\it (iii)} observe that if $g \in {\bf U}_*(\Omega \times \Sigma)$
then the relation (4.12) holds, and consequently the complex conjugation of
Eqs. (4.1) gives

$$ -i \hbar \partial_z \overline{E_{\lambda}} = (1 - \bar{\lambda}) a_z *
\overline{E_{\lambda}}, $$

$$ -i \hbar \partial_{\bar z} \overline{E_{\lambda}} = (1 - \bar{\lambda}^{-1})
a_{\bar z} * \overline{E_{\lambda}}.  \eqno(4.14)$$
By simple manipulations one quickly finds that the system (4.14) can be
equivalently rewritten as follows

$$ i \hbar \partial_z \overline{E_{\lambda}}^{ \ -\buildrel{*}\over{1}} = (1 -
\bar{\lambda})\overline{E_{\lambda}}^{ \ -\buildrel{*}\over{1}}*a_z, $$

$$ i \hbar \partial_{\bar z} \overline{E_{\lambda}}^{ \ -\buildrel{*}\over{1}}
= (1 - \bar{\lambda}^{-1})\overline{E_{\lambda}}^{ \
-\buildrel{*}\over{1}}*a_{\bar z}. \eqno(4.15)$$
Then, changing  $\lambda \leftrightarrow \bar{\lambda}^{-1}$ we get

$$ i \hbar \partial_z \overline{E_{\bar{\lambda}^{-1}}}^{- \
\buildrel{*}\over{1}} = (1 - {\lambda}^{-1})\overline{E_{\bar{\lambda}^{-1}}}^{
\ -\buildrel{*}\over{1}}* a_z, $$

$$ i \hbar \partial_{\bar z} \overline{E_{\bar{\lambda}^{-1}}}^{ \
-\buildrel{*}\over{1}} = (1 - \lambda)\overline{E_{\bar{\lambda}^{-1}}}^{
\ -\buildrel{*}\over{1}}*a_{\bar z}. \eqno(4.16)$$

Comparing (4.16) with (4.1), and employing  also the assumption of the
existence and the uniqueness of the solution to Eqs. (4.1) one obtains (4.13).
$\diamondsuit$

It is evident that Theorem 4.1 is a ``weak'' version of the Uhlenbeck's Theorem
2.2 as we {\it assume} and {\it don't prove} the existence and uniqueness of
solution to the Lax pair. In our case this problem is rather subtle and we have
not succeeded in solving it in all generality.

We now prove a theorem which corresponds exactly to Uhlenbeck's Theorem 2.3
[\Uhlen].

\subsection{Theorem 4.2}

Let the function $E = E(\lambda,x,y,p,q) \in C^{\infty}({\bf C}^* \times \Omega
\times \Sigma)$ be such that $E_1 = E(1,x,y,p,q) = f(p,q)$ and the expressions
${ E_{\lambda}^{-\buildrel{*}\over{1}} * \partial_{\bar z} E_{\lambda} \over 1-
\lambda}$ and ${ E_{\lambda}^{-\buildrel{*}\over{1}} * \partial_z E_{\lambda}
\over 1- {\lambda}^{-1}}$, $\lambda \in {\bf C}^*$, are independent of
$\lambda$. Then $g = E_{-1
}$ is a $*$-harmonic map from $\Omega$ into $C^{\infty}(\Sigma)$.

\noindent
{\it Proof:}

Define

$$a_{\bar z}:= {i \hbar  E_{\lambda}^{-\buildrel{*}\over{1}} * \partial_{\bar
z} E_{\lambda} \over 1- \lambda} \ \ {\rm and}       \ \ a_z := {i \hbar
E_{\lambda}^{-\buildrel{*}\over{1}} * \partial_z E_{\lambda} \over 1-
{\lambda}^{-1}}   \ \ \  \lambda \
in {\bf C}^*. \eqno(4.17)$$

By the assumption, $a_z$ and $a_{\bar z}$ are independent of $\lambda$. Thus,
in fact, we arrive at the Lax pair (4.1). Consequently, Eq. (4.3) is satisfied.
{}From (4.17) one quickly finds that

$$a_{\bar z}:= {i \hbar\over 2} E_{-1}^{-\buildrel{*}\over{1}} * \partial_{\bar
z} E_{-1}  \ \ {\rm and}       \ \ a_z := {i \hbar \over  2}
E_{-1}^{-\buildrel{*}\over{1}} * \partial_z E_{-1}. \eqno(4.18) $$
Inserting (4.18) into (4.3) we get the thesis. $\diamondsuit$

Now we are in a position to prove the fundamental theorem

\subsection{Theorem 4.3}

Let the function $E=E(\lambda,x,y,p,q) \in C^{\infty}({\bf C}^* \times \Omega
\times \Sigma)$ be as in the Theorem 4.2, and let $\Theta = \Theta(x,y,p,q)$
 be defined by
$$ \partial_{\bar z} \Theta = \hbar E_{-1}^{-\buildrel{*}\over{1}} *
\partial_{\bar z} E_{-1} \ \ \ {\rm and} \ \ \ \partial_z \Theta = - \hbar
E_{-1}^{-\buildrel{*}\over{1}} * \partial_z E_{-1}. \eqno(4.19)$$
Then $\Theta$ satisfies the Moyal deformation of Husain's heavenly equation,
(2.17) or (4.5). Moreover, if $E_{\lambda}$ fulfills the {\it $*$-unitary
condition} (4.13) then $\Theta$ can be chosen so as to be real.

\noindent
{\it Proof:}

Straightforward by the Theorem 4.2 and by the relations (4.4). $\diamondsuit$

{}From the previous considerations (see (2.28)) it follows that if the
function $\Theta = \Theta (x,y,p,q)$ defined by (4.19) is analytic in
$\hbar$ then $\Theta_0$ satisfies Husain's heavenly equation (2.29). Given a
$*$-harmonic map $g$ from $\Omega$ into $C^{\infty}(\Sigma)$ we define the {\it
extended solution corresponding to $g$} to be the solution
$E=E(\lambda,x,y,p,q) \in C^{\infty}({\
bf C}^* \times \Omega \times \Sigma)$ of the Lax pair (4.1) with $a_z$ and
$a_{\bar z}$ defined by (4.6).

Then by the analogy to Uhlenbeck's definition [\Uhlen] (see also [\Ward,\Ana])
one has

\subsection{Definition 4.1}

Let $g = g(x,y,p,q)$ be a $*$-harmonic form $\Omega$ into $C^{\infty}(\Sigma)$
such that it has an extended solution $E_{\lambda}$ of the form

$$ E_{\lambda} = \sum_{k = 0}^n T_k \  \lambda^k,  \ \ \ \ T_k \in
C^{\infty}(\Omega \times \Sigma), \eqno(4.20)$$
satisfaying the $*$-unitary condition (4.13). The self-dual metric (if it
exists) defined by $g$ is called the {\it n-gravitational uniton}.

The question of existence and of the interpretation of the gravitational
unitons will be considered in a separate paper. Here we have only shown that
the Moyal bracket algebra, enables one to consider self-dual gravity as the
principal chiral model.

\chapter{ The self-dual gravity action as the WZW-like action }

In this section we are going to show that the action leading to the
Moyal deformation of Husain's heavenly equation (2.17) can be written in
the WZW-like form within the Moyal bracket algebra formalism.
Consequently, the action for Husain's heavenly equation (2.29) appears
to be the $\hbar \to 0$ {\it limit of the WZW-like action in the Moyal
bracket algebra.}  To this end consider the SU$(N)$ chiral model. Let
$\Omega = S^2$ with the radius $r \to \infty$ and let $B$ be a three
dimensional manifold such that $\Omega$ is the boundary of $B$ {\it
i.e.,} $\partial B = \Omega$. The Cartesian coordinates on $\Omega$ will
be denoted by $\sigma^{\mu}$, $\mu = 1,2$. The coordinates on $B$ will
be denoted by $t^i$, $i=1,2,3.$ Then employing the formula (2.7)  which
defines the Lagrangian leading to the principal chiral equations (2.3)
one quickly finds that Eqs. (2.3) can be derived from the following  action

$$ I(\theta) = - {\alpha \over 2} \int_{\Omega} d^2 \sigma {\rm Tr} \big[
(\partial_{\mu} \theta )(\partial_{\mu} \theta ) \big] + {\alpha \over 3}
\int_B {\rm Tr} \big( d\theta \wedge d\theta \wedge d\theta \big)$$

$$  = - {\alpha \over 2} \int_{\Omega} d^2 \sigma {\rm Tr} \big[
(\partial_{\mu} \theta )(\partial_{\mu} \theta ) \big] + {\alpha \over 3}
\int_B  d^3 t \  \epsilon^{ijk} {\rm Tr} \big[(\partial_i \theta )(\partial_j
\theta )
(\partial_k \theta ) \big], \eqno(5.1)$$
where in the integral over $B$ the function $\theta$ is considered to be
extended over $B$. The action (5.1) evidently resembles the WZW action [9,10].

Analogously, keeping in mind the Lagrangian (2.24) one obtains the action for
the Moyal deformation of Husain's heavenly equation (2.17) to be

$$ I^{(M)}_{SG} (\Theta) ={1\over 2} \int_{\Omega \times \Sigma} d^2 \sigma dp
dq \big(\partial_{\mu} \Theta \big) * \big(\partial_{\mu} \Theta \big) - {1
\over 3i \hbar} \int_{B \times \Sigma} d^3t dp dq \epsilon^{ijk}
\big[\big(\partial_i \Theta \big)*\
big(\partial_j \Theta \big)*\big(\partial_k \Theta \big) \big]. \eqno(5.2)$$
Then we rewrite (5.2) in the following form

$$ I^{(M)}_{SG} (\Theta)= {1\over 2} \int_{\Omega \times \Sigma} d^2 \sigma dp
dq \big(\partial_{\mu} \Theta \big) * \big(\partial_{\mu} \Theta \big) - {1
\over 6} \int_{B \times \Sigma} d^3t dp dq \epsilon^{ijk} \big[\partial_i
\Theta * \{ \partial_j \Theta, \partial_k \Theta \}_M \big]. \eqno(5.3)$$

Consequently, taking the $\hbar \to 0$ limit of (5.3) one gets the WZW-like
action for Husain's heavenly equation to be

$$ I_{SG} (\Theta_0)= {1\over 2} \int_{\Omega \times \Sigma} d^2 \sigma dp dq
\big(\partial_{\mu} \Theta_0 \big) \big(\partial_{\mu} \Theta_0 \big) - {1
\over 6} \int_{B \times \Sigma} d^3t dp dq \epsilon^{ijk} \big[ \partial_i
\Theta_0  \{ \partial_j \Theta_0, \partial_k \Theta_0 \}_P \big]. \eqno(5.4)$$

The fundamental question is how one can interpret geometrically the
``Wess-Zumino'' term in (5.2). We intend to consider this question soon.

\vskip 2truecm

We are grateful to I.A.B. Strachan and T. Matos for their suggestions and also
to M. Dunajski for his interest in this paper. One of us (M.P.) thank the staff
of Departamento de F\'{\i}sica at CINVESTAV, M\'exico, D.F., for warm
hospitality. The work is s
upported by CINVESTAV, CONACyT and SNI, M\'exico, D.F., M\'exico.

\endpage
\refout
\end